# Spatial properties of the complex decameter type II burst observed on 31 May 2013

V.V. Dorovskyy [1], V.N. Melnik [1], A.A. Konovalenko [1], A.I. Brazhenko [2], H.O. Rucker [3]

[1] Institute of Radio Astronomy of NASU, Kharkov, Ukraine,
[2] Poltava Gravimetric Observatory of NASU, Poltava, Ukraine,
[3] Commission for Astronomy of AAS, Graz, Austria.

**Abstract.**

We present the results of observations of complex powerful type II burst associated with narrow Earth-directed CME, which was ejected at around 11 UT on 31 May 2013. The observations were performed by radio telescope UTR-2, which operated as local interferometer, providing the possibility of detection of the spatial parameters of the radio emission source. There are also polarization data from URAN-2 radio telescope.

The CME was detected by two space-born coronagraphs SOHO/LASCO/C2 and STEREO/COR1-BEHIND, and was absolutely invisible for STEREO-AHEAD spacecraft.

The associated type II burst consisted of two successive parts of quite different appearance on the dynamic spectrum. The first burst was narrow in frequency, had cloudy structure and was completely unpolarized while the second one represented rich herring-bone structure and exposed high degree of circular polarization. Both parts of the whole event reveal band splitting and well distinguished harmonic structure.

The positions and sizes of the sources of the type II burst were found using cross-correlation functions of interferometer bases.

The sources of the type II bursts elements were found to be of about 15 arcmin in size in average, with the smallest ones reaching as low as 10 arcmin. Corresponding brightness temperatures were estimated. In most cases these temperatures were between $10^{11}$ and $10^{12}$ K with maximum value as high as $10^{14}$ K.

The spatial displacement of the source was measured and model independent velocities of the type II burst sources were determined.

**Introduction**

Solar type II bursts are the manifestations of CME movement through the solar corona and originate from the shock formed ahead of CME. In its turn CME is one of the most effective drivers of the space weather changes. Though type II bursts are observed in wide frequency range the bursts detected at meter and decameter-hectometer wavelengths are believed to be connected with the potentially most geoeffective CMEs [*Gopalswamy, N, et.al*., 2005].

The geoeffectiveness of CME is mostly defined by its mass, velocity and direction. The velocity of CME can be detected in three ways: by coronagraph observations in visible light, by the frequency drift rate of type II bursts and by the speed of the type II source displacement. The first method gives correct results for CMEs which propagate close to the sky plane and is not applicable for Earth-directed CMEs (when observing from Earth). This problem can be solved by multi-spacecraft observations such as STEREO mission. In addition multi-spacecraft observations also allow to define the direction of CME movement. However the accuracy of this method depends on mutual positions of the STEREO satellites with respect to the Earth and these positions are changing with time. The second method is indirect and model dependent one. In addition it gives the radial component of the velocity vector and is inaccurate for CMEs which move at the substantial angle to the density gradient. This method is especially effective in cases of Earth-directed CMEs. The third method deals with real spatial displacement of the source of type II burst with time regardless the chosen coronal model.

It is known also that type II bursts reveal a wide variety of morphological features, such as fundamental and harmonic emission, band splitting, cloudy structure, herring-bone (HB) structure etc [*Nelson and Melrose*, 1985, *Melnik et al.,* 2004, *Dorovskyy, et al.,* 2015]. The interpretations of all these features are based on their spectral properties assuming some models of plasma distributions in the corona and the shock. For instance the band splitting is attributed to the emission from ahead and behind the shock as upstream and downstream emission [*Vršnak, et.al*., 2004], HB sub-bursts are manifestation of the electron acceleration towards and outwards the Sun [*Cairns and Robinson,* 1987]. Spatial measurements with high time, frequency and spatial resolution may confirm or call these assumptions into question.

Thus study of the spatial properties of the sources of type II bursts may improve the quality and reliability of the space weather diagnostics and will lead to better understanding of underlying physics as well.

**Observations**

Since 2011 observations of solar radio emission at the UTR-2 radio telescope are carried out in the interferometric mode which gives an opportunity for estimation of sizes and locations of solar radio bursts sources. The UTR-2 radio telescope having effective collecting area of up to 150000 m$^2$ consists of 12 separate sections which can be configured in different ways. For the discussed observations the sections were grouped into three separated in space antennas: North (N), West (W) and South (S) by 4 sections in each thus forming a local interferometer with two orthogonal baselines B1 and B2 each being 674 m long (see Figure 1). These two baselines are inclined by 45° to the meridian.

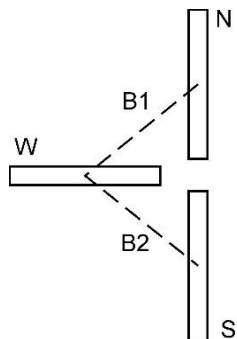

*Fig. 1 The UTR-2 radio telescope as two-base interferometer.*

As a backend two spectropolarimeters DSPZ [*Zakharenko, V., et.al.*, 2016] are used. These digital analyzers calculate complex correlation function in instantaneous frequency band 10-32 MHz for each of two bases. Then the absolute value of the correlation function carries information about the size of the source via visibility function while the phase of the correlation function is determined by the source angular displacement.

On 31 May 2013 complex powerful type II burst was observed by UTR-2, URAN-2 and NDA radio telescopes. The overall view of the burst is shown in Figure 2. The whole event started at 11:12 UT with the group of type III bursts. In this particular case these type III bursts are the precursor of following type II bursts and coincide in time with solar flare of class C1.1 above the active region NOAA11761 (S18E23). The type II burst itself consists of two parts separated in time. First part lasted from 11:22:30 till 11:25:30 UT and the second one – from 11:28 till 11:39 UT

Both parts of the burst exhibit band splitting. The lanes were separated in frequency by 4.5 MHz and 6 MHz respectively. The dynamic spectrum obtained by the NDA shows also the harmonic structure of the bursts (see Figure 3). The burst appeared to be bounded both above 65 MHz and below 10 MHz. The burst appeared to be extremely intense with maximum fluxes as high as 10000 s.f.u. at frequency around 20 (1 solar flux unit equals $10^{-22} Wm^2 Hz^{-1}$).

No sign of type II burst at lower frequencies was found by SWAVES receivers onboard STEREO spacecraft.

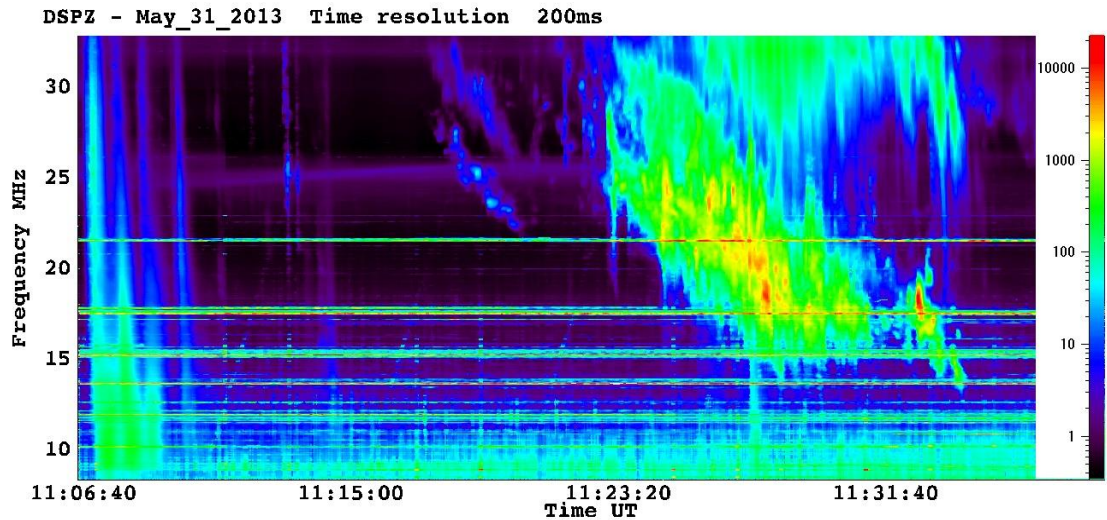

Fig. 2 *Type II burst recorded on 31 May 2013 by UTR-2 radio telescope.*

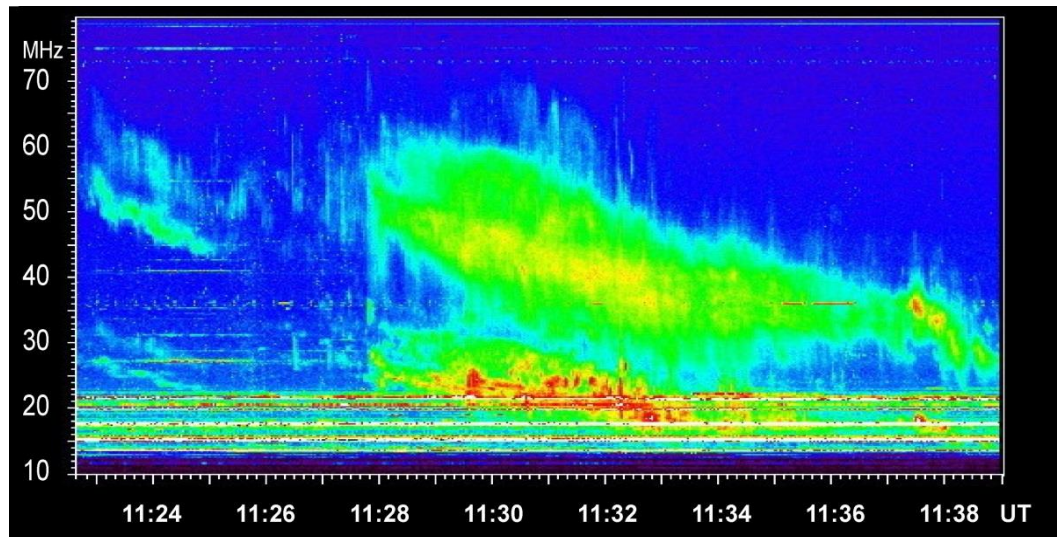

Fig. 3 *Type II burst recorded on 31 May 2013 by NDA radio telescope.*

This type II burst was apparently initiated by the CME recorded by coronagraphs SOHO/LASCO-C2 at 12:12 UT (https://cdaw.gsfc.nasa.gov/CME_list/) and by STEREO-B/COR1 at 11:20. At the same time this CME was not visible for STEREO-A spacecraft.

**Analysis and discussion**

*a) Determination of the direction and the speed of the CME propagation using optical data*

The views of CME from different locations and mutual positions of the satellites are shown in Figure 4. The CME of interest is marked with white arrows. The speeds of the CME measured from SOHO and STEREO-B spacecraft are 280 km·s$^{-1}$ and 340 km·s$^{-1}$ respectively. Taking into account the fact that this CME isn't visible for STEREO-A satellite we may conclude that this CME propagated at angle ~30° Eastward to the line of sight with the velocity of about 340 km·s$^{-1}$. In this case the CME moves practically in the sky plane for STEREO-B vehicle and is 30° projection for the SOHO spacecraft. For such a trajectory STEREO-A couldn't observe the CME because it was completely behind the Sun.

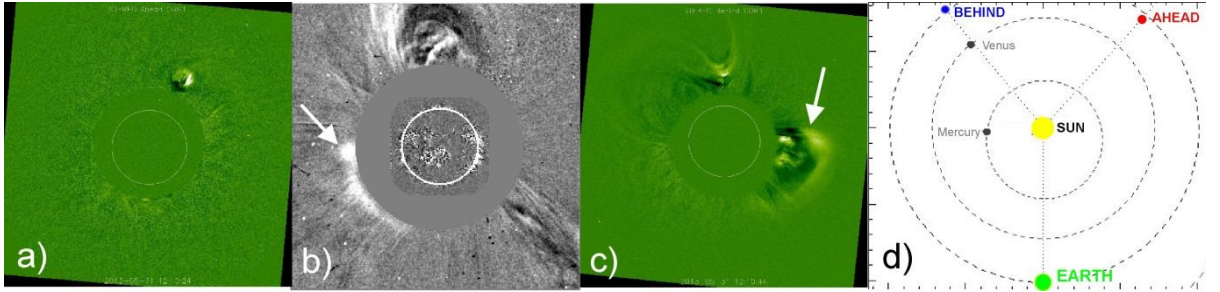

*Fig. 4 View of the CME at 12:12 UT from STEREO-A (a), SOHO (b) and STEREO-B (c) apparatus and the positions of the STEREO spacecraft on 31 May 2013 (d).*

*b) Determination of the CME speed via frequency drift rate of the type II bursts.*

It is well known that in frames of plasma emission mechanism the frequency drift rate of any burst depends on the corona density profile and the source velocity

For current analysis we apply the Newkirk (1961) corona model. For the fundamental component of the discussed burst the drift rate is not constant for the burst lifetime. In the beginning the drift rate was about -40 km·s$^{-1}$, that corresponded to the radial velocity of about 800 km·s$^{-1}$. The tail of the type II bursts has practically no drift. The radial velocity of the first part of the bursts is more than twice higher than that of the CME speed according to coronagraph data. The possible reason of such a discrepancy is discussed below.

*c) Using spatial properties of the type II burst for retrieving CME parameters*

The possibility of estimation of the radio sources sizes and positions using the UTR-2 radio telescope as an interferometer was shown in [Shepelev, V.A., 2015]**.**

The size of radio source in this case is determined through the visibility function [*Thompson, A.R., et.al.*, 1986]

$$\gamma = \exp[-\left(\frac{\pi \theta L}{2\lambda\sqrt{ln2}}\right)^2], \qquad (1)$$

where $\theta$ is the half-maximum angular size of a source, $L$ is the base length of the interferometer and $\lambda$ is the wavelength of the radio emission. Note that equation (1) is only valid for sources with Gaussian brightness distribution. In the case when the interferometer is composed of the element of the same array (providing identical antenna parameters) the visibility function can be calculated as [Shepelev, V.A., 2015]:

$$\gamma = \frac{P_c}{\sqrt{P_1 \cdot P_2}}, \qquad (2)$$

where $P_c$ is the absolute value of the correlation function, and $P_1$ and $P_2$ are the power received by two antennas of the interferometer base. Using equations (1) and (2) it is possible to get the source size along the direction of the baseline.

In its turn the phase of the correlation function allows to define angular deflection of a source from the interferometer axis.

The spatial properties of the first part of the type II burst, which manifests well distinguished band splitting are shown in Figure **5.**

We have analyzed spatial properties of two lanes of one type II burst taken at one moment of time. The source of higher frequency lane is drawn in red while lower frequency lane is marked with blue color. Their sizes are almost equal (14` and 16` respectively) and this gives us the effective brightness temperatures of these sources of $2 \cdot 10^{12}$K and $4 \cdot 10^{11}$K. The obtained sizes of the type II bursts sources appeared to be slightly smaller than the sizes of type III bursts, measured by the same radio telescope with the similar method at the same frequencies [Melnik, et al., 2017].

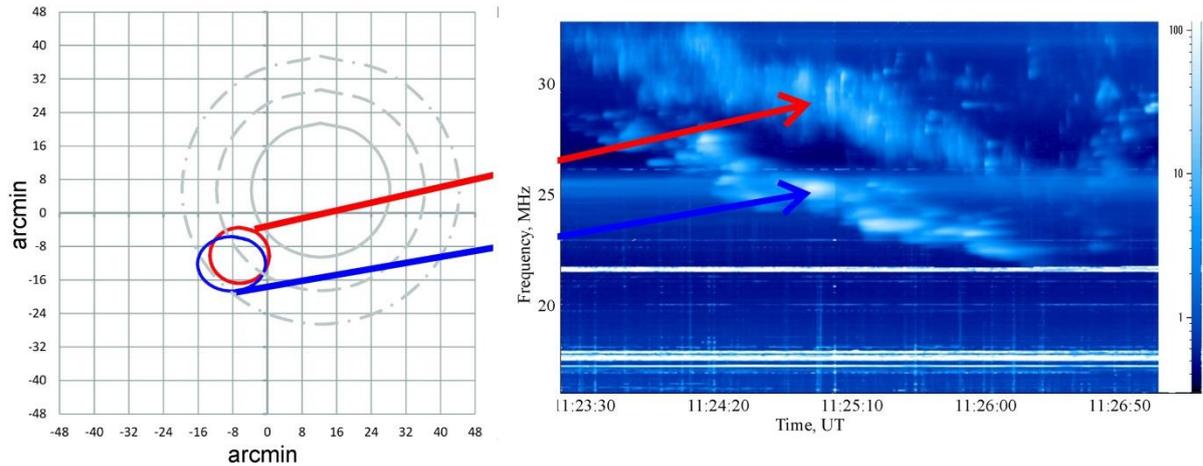

*Fig. 5 Positions and sizes of the sources of type II burst with band splitting (a) and corresponding dynamic spectrum (b).*

It is also evident that higher frequency lane is situated closer to the Sun than lower frequency lane. In addition we must note that these two lanes have different morphology as can be easily seen in Figure 5b. It is evident that the higher frequency lane consists of "clouds" extended along the frequency axis (vertically) while the lower frequency lane is composed of exclusively horizontally stretched structures. These facts show that the sources of two lanes are different and thus speak in favor of upstream and downstream model of band splitting of type IIs.

In the same way we have analyzed how the source of one individual lane moves in space with time. Apparently the source moves outwards the Sun (Figure 6). The velocity obtained from this displacement is around 900 km·s$^{-1}$.

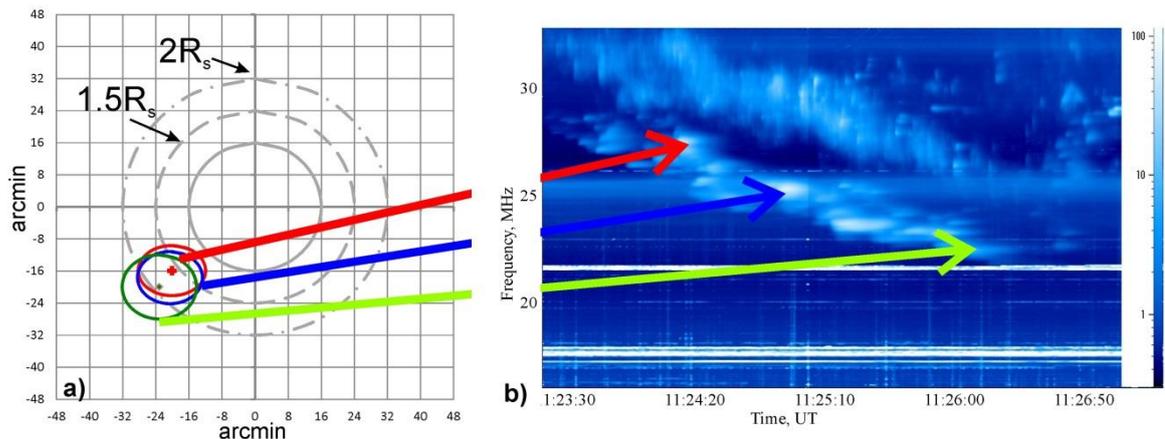

**Fig. 6 Displacement of the type II burst source with time (a) and its spectrum (b)**

This velocity is very close to that obtained from the frequency drift rate and much higher than velocity provided by coronagraphs. To our opinion this discrepancy may possibly be connected with migration of the radio source across the shock surface. Indeed the obtained size of the radio source is much less than the size of CME itself and thus the source may change its position at the shock with time causing apparent increase of its velocity.

The formation of the HB structure is generally attributed to the electron beams acceleration at the shock in two opposite directions [*Cairns and Robinson,* 1987]. Figures 7 and 8 evidently illustrate the inward and outward movement of the sources of HB sub-bursts with opposite frequency drift signs. The velocities of these movements are about $0.3c$, and are close to the type III electrons velocity.

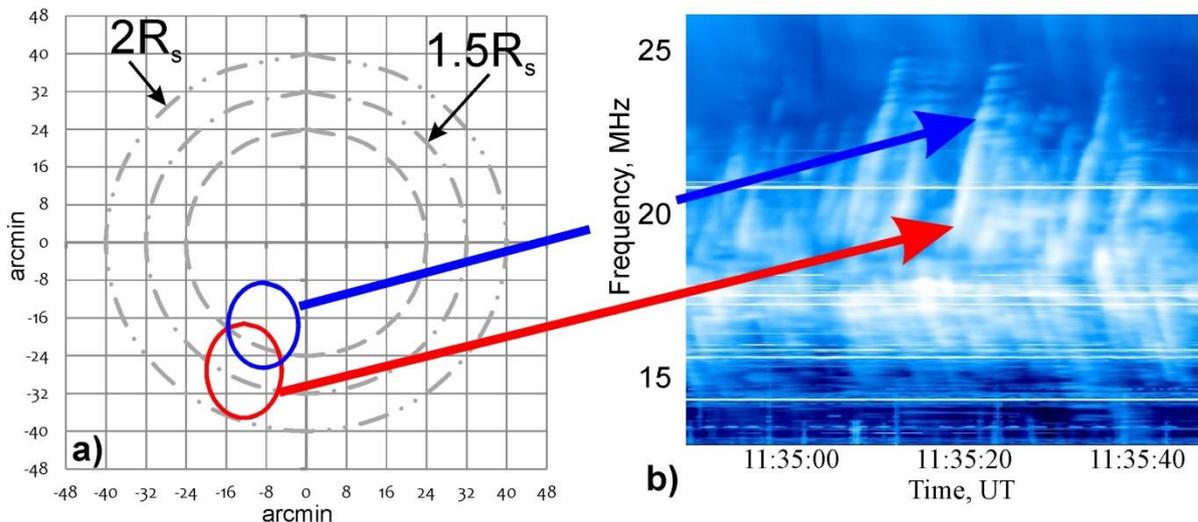

Fig. 7 *Displacement of the reverse HB sub-bursts source with time (a) and its spectrum (b).*

### d) Polarization properties of the type II burst.

Data from URAN-2 radio telescope allow to measure the polarization of radio emission. For this particular burst the fundamental HB sub-bursts reveal high degree of circular polarization reaching as high as 70%. The fundamental counterparts are weakly polarized. This is in good agreement with theoretical view. On the contrary both the fundamental and the harmonic components of the first part of the type II burst (without HB structure) are completely unpolarized. This may indicate that the sources of different parts of type II burst are located at different sites with different conditions.

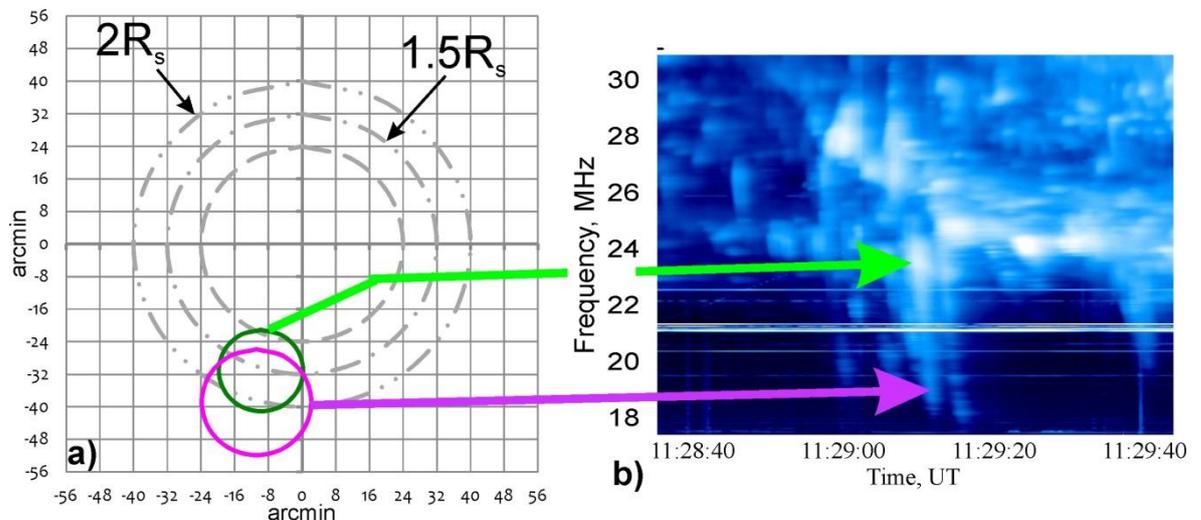

**Fig. 8 Displacement of the reverse HB sub-bursts source with time and its spectrum (b)**

**Conclusions**

Present work has shown the possibility of detecting of spatial properties of fine structured solar radio bursts, such as type II bursts with band splitting and HB structure. The speeds of sources of different parts of the complex type II burst were obtained by two independent methods: by the frequency drift rate (model dependent value) and by the source spatial displacement (model independent value). The obtained velocities, 800 km/s and 900 km/s respectively, appeared to be very close. At the same time they both differed considerably from the associated CME speed, obtained from coronagraphs data. This effect could be possibly connected with migration of the radio burst source across the surface of the shock.

The sizes of the type II burst appeared to be much smaller than the apparent sizes of the shock with which they are associated. They are also slightly smaller than normal type III bursts sources. High flux densities and small sizes result in extremely high brightness temperatures of the type II burst source, which in few cases reach the value of about $10^{14}$K.

Obtained locations of the bursts sources indicate that band splitting is likely the result of emission from ahead and behind the shock front. They also show that HB structure is most probably the result of electrons acceleration towards and outwards the Sun from the shock.

The obtained spatial properties of the type II burst sources also point out that different fragments of the type II burst of the same frequency may originate from different parts of the shock surface. This fact may lead to uncertainty in the source speed measurements using frequency drift rates of type II bursts.

**Acknowledgment**

The work was partially performed under the support of the European FP-7 project SOLSPANET (F9P7-People-2010-IRSES-269299).